\documentclass[aps,twocolumn,showpacs]{revtex4}
\usepackage{amssymb}
\usepackage{amsmath}
\usepackage{amscd}
\usepackage{graphicx}
\usepackage{subfigure}
\usepackage{float}
\usepackage{array}
\begin{document}

\title[Short Title]{Quantum phase transition of polaritonic
excitations in a multi-excitation coupled array}

\author{Li-Tuo Shen$^{1}$}
\author{Rong-Xin Chen$^{1}$}
\author{Huai-Zhi Wu$^{1}$}
\author{Zhen-Biao Yang$^{1}$}
\author{E. K. Irish$^{2}$}
\author{Shi-Biao Zheng$^{1}$}
\email{sbzheng11@163.com}

\affiliation{$^{1}$Lab of Quantum Optics, Department of Physics,
Fuzhou University, Fuzhou 350002, China\\ $^{2}$SUPA, School of
Physics and Astronomy, University of St. Andrews, St. Andrews, KY16
9SS, United Kingdom}

\begin{abstract}

We analyze the quantum phase transition-like behavior in the lowest
energy state of a two-site coupled atom-cavity system, where each
cavity contains one atom but the total excitation number is not
limited to two. Utilizing the variance of the total excitation
number to distinguish the insulator and superfluid states, and the
variance of the atomic excitation number to identify the polaritonic
characteristics of these states, we find that the total excitation
number plays a significant role in the lowest-energy-state phase
transitions. In both the small hopping regime and the small
atom-field interaction regime, we identify an interesting coexisting
phase involving characteristics of both photonic superfluid and
atomic insulator. For small hopping, we find that the signature of
the photonic superfluid state becomes more pronounced with the
increase in total excitation number, and that the boundaries of the
various phases shift with respect to the case of $N=2$. In the limit
of small atom-field interaction, the polaritonic superfluid region
becomes broader as the total excitation number increases. We
demonstrate that the variance of the total excitation number in a
single site has a linear dependence on the total excitation number
in the large-detuning limit.

\end{abstract}

\pacs{42.50.Pq, 05.70.Fh, 03.67.Lx}
  \keywords{polaritonic excitation, insulator-superfluid transition, quantum phase transition, multi-excitation coupled array}
\maketitle

\noindent

\section{Introduction}

Strongly correlated phenomena in controllable quantum many-body
systems have attracted great attention in optical lattices
\cite{Advphys-56-243-2007,RMP-80-885-2008} and Josephson-junction
arrays \cite{PhyRep-355-235-2001}. One of the simplest and most
important models describing these light-matter interactions is the
Jaynes-Cummings-Hubbard (JCH) model
\cite{Nature-2-856-2006,Nature-2-849-2006,PRA-76-031805-2007}, which
describes a coupled array of cavities each containing a two-level
system. Similar to the Bose-Hubbard (BH) model
\cite{PRL-69-2971-1992,PRL-81-3108-1997} used for cold atoms in an
optical lattice, the JCH model exhibits the photon blockade effect
\cite{PRL-79-1467-1997,Nature-436-87-2005} and the
Mott-insulator-to-superfluid quantum phase transition
\cite{Nature-415-39-2002}.

Coupled cavity-QED arrays can operate at high temperatures and allow
for individual site addressing. These features, together with
progress in realizing the strong light-matter coupling regime in
both atomic and solid-state systems
\cite{RMP-73-565-2001,Nature-445-896-2007}, are attracting more and
more attention to the quantum phase transitions in coupled-cavity
arrays captured by the JCH model. These quantum phase transitions
are due to the transfer of excitations from polaritonic states to
photonic states, where polaritons are superpositions of photons and
excitations of the atoms or atom-like structures, rather than purely
bosonic or purely fermionic excitations. Most previous studies
related to quantum phase transitions in coupled-cavity arrays have
focused on the large site number and large atom number limits
\cite{PRL-99-186401-2007,PRB-82-045126-2010,PRA-77-022103-2008,PRA-84-033817-2011},
which are analogous to the purely bosonic BH model and can be
analytically solved within the mean-field approximation.

Recent research shows evidence that novel quantum phase
transition-like behavior may appear in finite systems involving a
very few interacting sites and a small number of two-level systems
\cite{PRA-77-033801-2008,PRA-80-043825-2009,
PRA-77-023620-2008,PRA-84-063816-2011,PRA-80-060301-2009}, where the
transition behavior becomes dependent on the number of sites and
two-level systems. Greentree \emph{et al.} \cite{Nature-2-856-2006}
showed that the Mott-insulator to superfluid quantum phase
transition could appear in a mesoscopic two-dimensional coupled
array. Hartmann \emph{et al.} \cite{Nature-2-849-2006} used a
four-level atom to simulate the effective on-site potential and the
Mott-insulator to superfluid phase transition. Angelakis \emph{et
al.} \cite{PRA-76-031805-2007} considered the simulation of an $XY$
spin model based on the Mott regime in a linear array of cavities,
each containing a two-level atom and a photon. Irish \emph{et al.}
\cite{PRA-77-033801-2008} demonstrated phase transitions of
polaritonic excitations in a two-site, two-excitation coupled array.
In a very recent ion-trap experiment \cite{PRL-111-160501-2013},
Toyoda \emph{et al.} reported the simulation of the quantum
transition of polaritonic excitations in a JCH model using two
trapped ions and phonons within the two-excitation Hilbert subspace.
Previous works on the two-site coupled array model were limited to
the two-excitation Hilbert subspace without considering the
situation with higher excitation numbers in the photonic states. It
is known that the number of photons plays an important role in the
coefficient of on-site repulsion for small finite systems
\cite{Nature-2-856-2006}, but the influence of extra photons on the
quantum phase transition in the lowest energy state of the coupled
atom-cavity system is still unclear.

In this work, we investigate quantum phase transition-like behavior
in the lowest energy state (within the $N$-excitation Hilbert
subspace) of a two-site coupled atom-cavity system, where each
cavity contains one atom but the total excitation number $N$ is not
limited to two. (Note that the descriptions `` insulator '' and
``superfluid '' used in this paper represent the localized and
delocalized states in the small finite system we consider. Such an
investigation can also be generalized to larger arrays, in which the
localization-delocalization transitions studied
 here approach genuine quantum phase transitions.) By not restricting the
 system to the two-excitation subspace, the requirement for cooling the
 cavity field in experimental realizations is loosened.
Our work is also applicable to ion-trap setups
\cite{PRL-111-160501-2013}, where the photons are replaced by
phonons. Section II establishes the model and its Hamiltonian. In
Sec. III we carry out an extensive analysis, both analytically and
numerically, of the lowest-energy-state properties for total
excitation number $N=4$. The limits of small hopping and of small
atom-field interaction are both considered in detail. In Sec. IV the
analysis is generalized to the case of higher excitation numbers. We
briefly conclude in Sec. V.

\section{System Hamiltonian}

We consider the system consisting of two sites, each supporting a
field mode and containing a single atom. Photons are able to hop
between these two field modes. Under the rotating-wave
approximation, our system is governed by the following Hamiltonian
($\hbar=1$):
\begin{eqnarray}\label{e1}
H&=&\sum_{j=1,2}[
w_{c}a^{\dagger}_{j}a_{j}+w_{a}|e_{j}\rangle\langle
e_{j}|+\lambda(a^{\dagger}_{j}|g_{j}\rangle\langle
e_{j}|\cr\cr&&+a_{j}|e_{j}\rangle\langle g_{j}|)]
+h(a^{\dagger}_{1}a_{2}+a_{1}a^{\dagger}_{2}),
\end{eqnarray}
where $a^{\dagger}_{j}$ and $a_{j}$ are the creation and
annihilation operators of the $j$th field mode with frequency
$w_{c}$. $|e_{j}\rangle$ and $|g_{j}\rangle$ represent
the excited and ground states of the $j$th atom with frequency
$w_{a}$. $\lambda$ is the atom-field coupling strength
and $h$ is the strength of the hopping between the two cavity fields. Note that
the excitation number of the total system is conserved since the
excitation number operator
$\hat{N}=\sum_{j=1,2}(|e_{j}\rangle\langle
e_{j}|+a^{\dagger}_{j}a_{j})$ commutes with the Hamiltonian $H$.

\section{Total excitation number $N=4$}

\subsection{Small hopping}
\begin{figure}
\center
  \includegraphics[width=0.9\columnwidth]{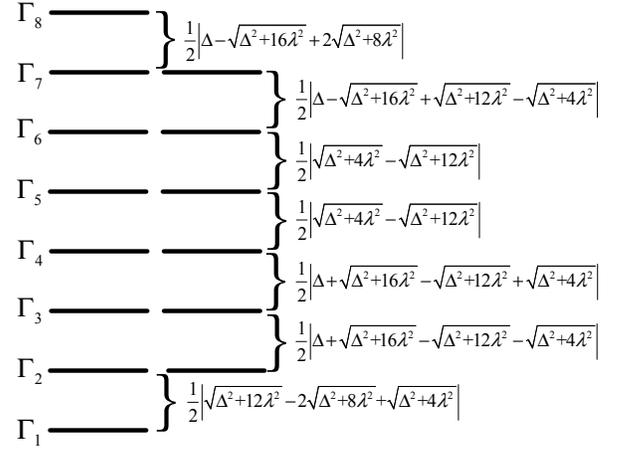} \caption{
  Energy-level difference between two nearest neighbor
  subspaces when there is no photon hopping and $N=4$.}\label{Fig.1.}
\end{figure}
When there is no photon hopping between the two sites, the
eigenstates in each site are given by the polaritonic states
\cite{PRA-77-033801-2008} :
\begin{eqnarray}
|0_{j}\rangle&=&|g_{j}\rangle|0_{j}\rangle,\label{e2}\\ \cr
|n^{-}_{j}\rangle&=&\sin(\frac{\theta_n}{2})|e_{j}\rangle|(n-1)_{j}\rangle-\cos(\frac{\theta_n}{2})|g_{j}\rangle|n_{j}\rangle,\label{e3}\\
\cr
|n^{+}_{j}\rangle&=&\cos(\frac{\theta_n}{2})|e_{j}\rangle|(n-1)_{j}\rangle+\sin(\frac{\theta_n}{2})|g_{j}\rangle|n_{j}\rangle,\label{e4}
\end{eqnarray}
where $j=1,2$. $|n_{j}\rangle$ ($n=1,2,3,...$) represents the
Fock state of the $j$th field mode, and
$\tan(\theta_{n})=2\lambda\sqrt{n}/\Delta$, where the detuning $\Delta=w_a-w_c$.
The corresponding energies for these eigenstates are:
\begin{eqnarray}
E_{j}^{0}&=&0,\label{e5}\\ \cr
E_{j}^{n^{-}}&=&nw_c+\frac{\Delta}{2}-\frac{1}{2}\sqrt{\Delta^2+4n\lambda^2},\label{e6}\\
\cr
E_{j}^{n^{+}}&=&nw_c+\frac{\Delta}{2}+\frac{1}{2}\sqrt{\Delta^2+4n\lambda^2}.\label{e7}
\end{eqnarray}
Unlike the previous study in Refs.
\cite{PRA-77-033801-2008,PRL-111-160501-2013} where the analysis is
restricted to the Hilbert subspace with only two excitations, our
analysis here focuses on the insulator-superfluid quantum phase
transition in the multi-excitation Hilbert space, i.e., the total
excitation number $N$ can be larger than two. For simplicity, we
consider the total excitation number to be an even number $N$, for
which the lowest energy state is nondegenerate. In the following, we
begin the analysis of quantum phase transition in the lowest energy
state of our atom-cavity system for the case $N=4$ and later
generalize it to $N>4$.

When $h=0$ and $N=4$, the eigenstates of the system are,
arranged in order of increasing energy,
\begin{eqnarray}
\Gamma_1&=&\{ |2_{1}^{-}\rangle\otimes|2_{2}^{-}\rangle \},\label{e8}\\
\cr
\Gamma_2&=&\{ |1_{1}^{-}\rangle\otimes|3_{2}^{-}\rangle, |3_{1}^{-}\rangle\otimes|1_{2}^{-}\rangle \},\label{e9}\\
\cr
\Gamma_3&=&\{ |0_{1}\rangle\otimes|4_{2}^{-}\rangle, |4_{1}^{-}\rangle\otimes|0_{2}\rangle \},\label{e10}\\
\cr
\Gamma_4&=&\{ |1_{1}^{+}\rangle\otimes|3_{2}^{-}\rangle, |3_{1}^{-}\rangle\otimes|1_{2}^{+}\rangle \},\label{e11}\\
\cr
\Gamma_5&=&\{ |2_{1}^{+}\rangle\otimes|2_{2}^{-}\rangle, |2_{1}^{-}\rangle\otimes|2_{2}^{+}\rangle \},\label{e12}\\
\cr
\Gamma_6&=&\{ |3_{1}^{+}\rangle\otimes|1_{2}^{-}\rangle, |1_{1}^{-}\rangle\otimes|3_{2}^{+}\rangle \},\label{e13}\\
\cr
\Gamma_7&=&\{ |4_{1}^{+}\rangle\otimes|0_{2}\rangle, |0_{1}\rangle\otimes|4_{2}^{+}\rangle \},\label{e14}\\
\cr
\Gamma_8&=&\{ |2_{1}^{+}\rangle\otimes|2_{2}^{+}\rangle
\}.\label{e15}
\end{eqnarray}
Figure 1 shows the energy differences between consecutive energy levels.

Note that the energetic ordering of the subspaces
$\Gamma_1\rightarrow\Gamma_8$ is independent of the parameters
$\lambda$ and $\Delta$. Notably, the gap between the lowest two
energy levels is $E_{\Delta_{1,2}}$ $=$
$\frac{1}{2}|\sqrt{\Delta^2+12\lambda^2}-
2\sqrt{\Delta^2+8\lambda^2}+\sqrt{\Delta^2+4\lambda^2}|$, which
approaches zero in the limits of both large positive and large
negative detuning. This is contrary to the situation with total
excitation number $N=2$, in which the energy difference between
different subspaces goes to infinity in the limit of large negative
detuning \cite{PRA-77-033801-2008,PRL-111-160501-2013}.

In order to distinguish the insulator and superfluid states in the
lowest energy state of the system, we use the variance of the total
excitation number on the first site $\hat{N}_1$ as a measure
\cite{PRA-76-031805-2007}:
\begin{eqnarray}
\Delta N_1=\langle \hat{N}_{1}^{2}\rangle-\langle
\hat{N}_{1}\rangle^{2},\label{e16}
\end{eqnarray}
where $\hat{N}_1=a_{1}^{\dagger}a_{1}+|e_{1}\rangle\langle e_{1}|$. A plot of $\Delta
N_1$ as a function of the detuning $\Delta$ and the photon
hopping strength $h$ is given in Fig. 2.
\begin{figure}
\center
  \includegraphics[width=1\columnwidth]{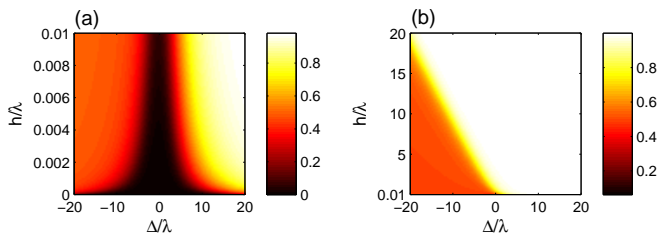} \caption{(Color
  online) [(a)-(b)] $\Delta N_{1}$ as a function of $\Delta$ and
  $h$ for the lowest energy state of the system when $N=4$.}\label{Fig.2.}
\end{figure}

For $\Delta=0$ and $h/\lambda<<1$, the atom-field interaction on one
site shifts the frequency of the field, causing a photon blockade
effect \cite{PRL-79-1467-1997,Nature-436-87-2005} that leads to a
large energy gap between the lowest two subspaces ($\Gamma_1$ and
$\Gamma_2$) and prevents additional photons from entering the site.
For $N=2$ \cite{PRA-77-033801-2008} the gap between the lowest two
energy levels is given by $(2-\sqrt{2})\lambda\simeq0.59\lambda$.
For $N=4$ the corresponding gap is
$(2\sqrt{2}-1-\sqrt{3})\lambda\simeq0.096\lambda$, indicating that
the hopping strength $h$ needed to overcome the photon blockade is
much smaller as compared with the case with $N=2$. For $N=4$, the
lowest energy state of the system is approximated by
$|2_{1}^{-}\rangle\otimes|2_{2}^{-}\rangle$, as shown in Fig. 3(a),
where $N_A$ is defined as the total excitation number of both atoms.
This lowest energy state contains two excitations on each site and
the state $|2_{j}^{-}\rangle$ is a maximally entangled state of the
atom and the field on the $j$th site. Fig. 3(b) shows that in the
lowest energy state the components with both atoms in the ground
state and both in the excited state are equally populated, which
corresponds to a polaritonic insulator state for exact resonance and
small hopping \cite{PRA-77-033801-2008}.

\begin{figure}
\center
  \includegraphics[width=1\columnwidth]{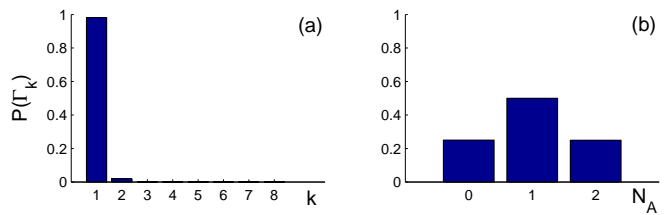} \caption{(Color
  online) (a) Probability distribution $P(\Gamma_{k})$ [$k=1,2,...,8,$
  corresponding to Eqs. (8)-(15)]
  of the lowest energy state with respect to the eigenspace of the
  Hamiltonian without hopping. (b) Probability distribution of the
  total atomic excitation number $N_A$.
  Parameters for both subfigures are $\Delta=0$ and $h=\lambda/200$.
   }\label{Fig.3.}
\end{figure}

When $h=0$, the energy gap $E_{\Delta_{1,2}}$ goes to zero in the
limit $\Delta/\lambda\rightarrow\pm\infty$, meaning the lowest level
of the system is degenerate. However, a small but nonzero hopping
value ($h/\lambda<<1)$ breaks this degeneracy, leading to a unique
state with the lowest energy. This lowest energy state involves a
superposition of polaritonic states, as seen in Fig. 4(a) and (c).
The eigenstates of each independent site vary with the sign of the
detuning, as seen in Fig. 4(b) and (d). For
$\Delta/\lambda\rightarrow\infty$, the lowest energy state is
approximately $|g_1g_2\rangle$ $\otimes$ $[$ $\frac{\sqrt{10}}{5}$
$|2_12_2\rangle$ $+$ $\frac{1}{2}$ $($ $|1_13_2\rangle$ +
$|3_11_2\rangle$ $)$ $-$ $\frac{\sqrt{5}}{10}$ $($ $|4_10_2\rangle$
+ $|0_14_2\rangle$ $)$ $]$, which is a delocalized photon state
(photonic superfluid state). For $\Delta/\lambda\rightarrow-\infty$,
the lowest energy state approximates $|e_1e_2\rangle$ $\otimes$ $[$
$\frac{\sqrt{2}}{2}$ $|1_11_2\rangle$ $+$ $\frac{1}{2}$ $($
$|0_12_2\rangle$ + $|2_10_2\rangle$ $)$ $]$, which is a coexisting
state with the characteristics of both photonic superfluid and
atomic insulator. This near-unity photon number is very different
from the system with total excitation number $N=2$ investigated in
Ref. \cite{PRA-77-033801-2008} in which the lowest energy state is
an atomic insulator state for small hopping and large negative
detuning. This result has a simple physical explanation. When
$\Delta$ is negative, the energy of the atomic excitation is lower
than that of the photon. To ensure the energy to be minimum, two
excitations should first be occupied by the atoms. The remaining two
excitations are populated in the photonic modes, with the
distribution being determined by the competition of the non-linear
Kerr effect induced by the dispersive atom-cavity interaction and
photon hopping.

\begin{figure}
\center
  \includegraphics[width=1\columnwidth]{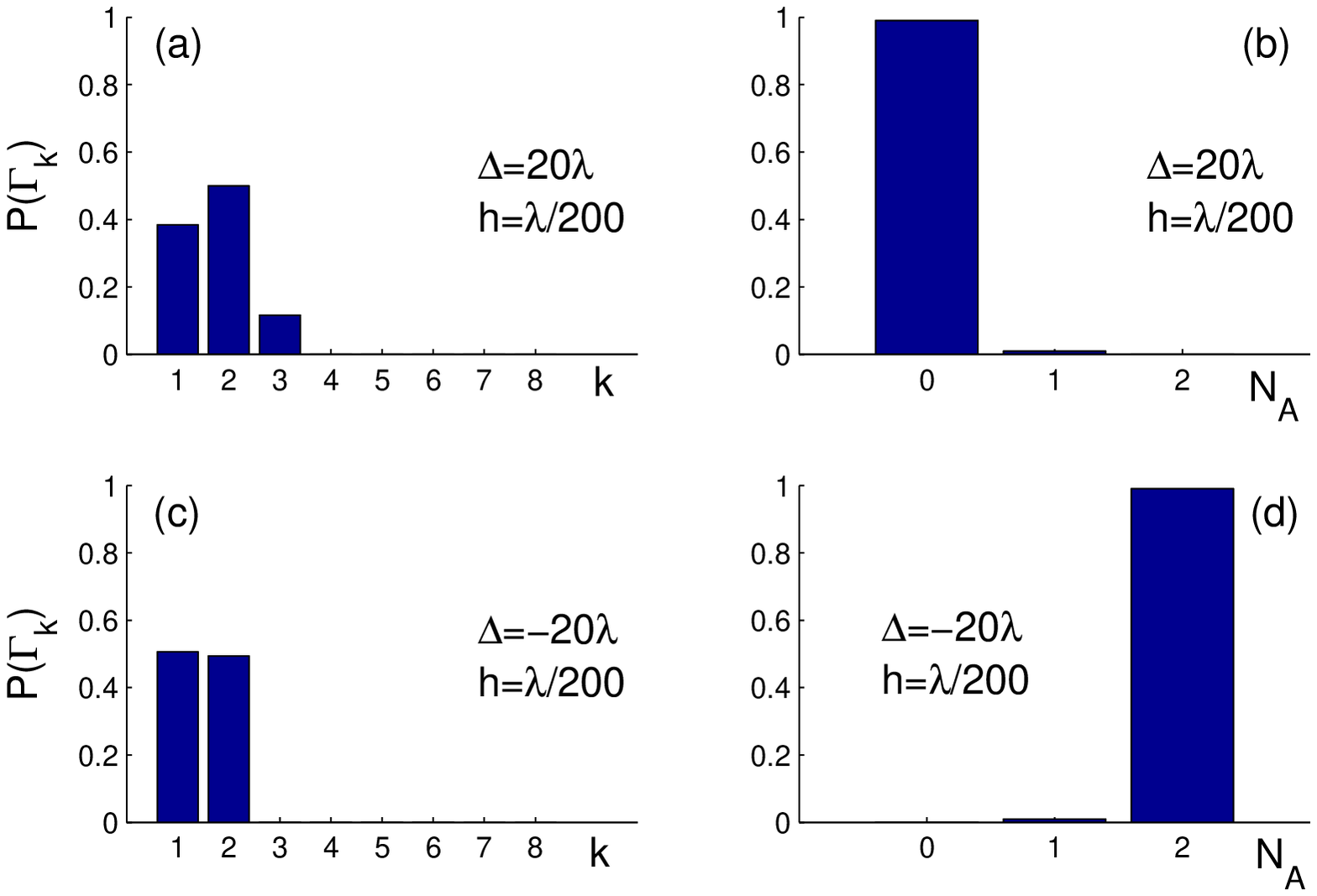} \caption{(Color
  online) [(a)-(c)] Probability distribution $P(\Gamma_{k})$ [$k=1,2,...,8,$
  corresponding to Eqs. (8)-(15)]
  of the lowest energy state with respect to the eigenspace of the
  Hamiltonian without hopping. [(b)-(d)] Probability distribution of the
  total atomic excitation number $N_A$.
   }\label{Fig.4.}
\end{figure}

Therefore, when the hopping is small, the lowest energy state of the
system undergoes quantum phase transitions from a polaritonic
insulator state near exact resonance to a photonic superfluid state
at large positive detuning.

In order to identify the polaritonic superfluid phase in the lowest
energy state, we take the variance of the excitation number of the
first atom $\Delta N_{1A}$ as a measure, where
$\Delta\hat{N}_{1A}=|e_1\rangle\langle e_1|$; the results are
plotted in Fig. 5(a) and (c). $\Delta N_{1A}$ is zero for the atomic
insulator state, but is nonzero for a state with polaritonic
characteristics. There are two regions with $\Delta N_{1A}=0$ in
Fig. 5(a): The region with $-\Delta>>\lambda$ is a coexisting state
where two excitations are occupied by the atoms and localized at
different sites, while the region with $\Delta>>\lambda$ is the
photonic superfluid state where no atom is excited.

The product $\Delta N_1\Delta N_{1A}$ may be used to characterize
the polaritonic superfluid state as shown in Fig. 5(b), (c), and
(d). $\Delta N_1\Delta N_{1A}$ is zero for the polaritonic insulator
state, but is nonzero for the polaritonic superfluid state. It is
apparent that the polaritonic superfluid state appears in the
near-resonance region for small hopping.

\begin{figure}
\center
  \includegraphics[width=1\columnwidth]{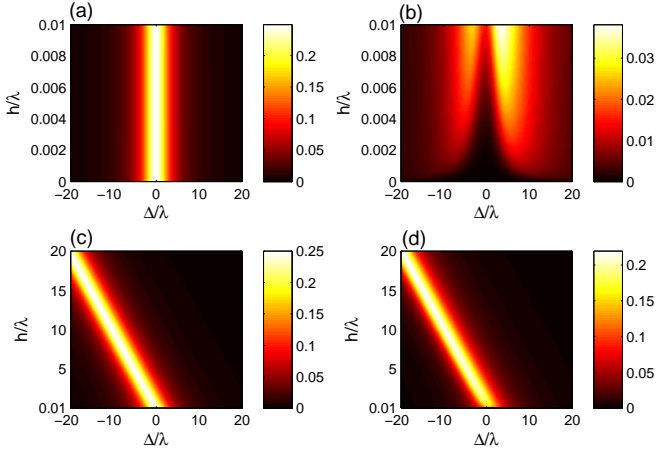} \caption{(Color
  online) [(a)-(c)] $\Delta N_{1A}$ as a function of $\Delta$ and $h$ for the
  lowest energy state of the system when $N=4$. [(b)-(d)] Product $\Delta
  N_{1}\Delta N_{1A}$ as a function of $\Delta$ and $h$ for the lowest energy
  state of the system when $N=4$.
   }\label{Fig.5.}
\end{figure}

\subsection{Small atom-field interaction}

In this section, the atom-field interaction in $H$ is taken as a
perturbation in order to analyze the system in the large-hopping
regime.

When $h<-\Delta$, the lowest eigenenergy is $4w_c+2\Delta-2h$, and
the corresponding eigenstate is the coexisting state
$|\varphi_{co}\rangle$ $=$ $|e_1e_2\rangle$ $\otimes$ $[$
$\frac{\sqrt{2}}{2}$ $|1_11_2\rangle$ $-$ $\frac{1}{2}$ $($
$|0_12_2\rangle$ + $|2_10_2\rangle$ $)$ $]$. This coexisting state
is similar to that in the regime $\Delta/\lambda\rightarrow-\infty$
for small hopping investigated in Sec. II.A., but is very different
from the case of $N=2$ discussed in Ref. \cite{PRA-77-033801-2008}
in which the lowest energy state is still an atomic insulator state
for large negative detuning. For $h>-\Delta$ the lowest eigenenergy
is $4w_c-4h$, and the corresponding eigenstate is the photonic
superfluid state $|\varphi_{ps}\rangle$ $=$ $|g_1g_2\rangle$
$\otimes$ $[$ $\frac{1}{4}$ $($ $|4_{1}0_{2}\rangle +
|0_{1}4_{2}\rangle$ $)$ $-$ $\frac{1}{2}$ $($ $|1_{1}3_{2}\rangle$ +
$|3_{1}1_{2}\rangle$ $)$ $+$ $\frac{\sqrt{6}}{4}$
$|2_{1}2_{2}\rangle$ $]$.

When $h=-\Delta$ and $\lambda=0$, the lowest level of energy
$4w_c-2h$ exhibits four-fold degeneracy with the eigenstates
$|\varphi_{co}\rangle$, $|\varphi_{ps}\rangle$,
$|\varphi_{1}\rangle$, and $|\varphi_{2}\rangle$, where
\begin{eqnarray}
|\varphi_{1}\rangle&=&\frac{2\sqrt{2}}{3}|e_{1}g_{2}\rangle\otimes[(|3_{1}0_{2}\rangle-|0_{1}3_{2}\rangle)
\cr&&-\sqrt{3}(|2_{1}1_{2}\rangle-|1_{1}2_{2}\rangle)],\label{e20}\\
|\varphi_{2}\rangle&=&\frac{2\sqrt{2}}{3}|g_{1}e_{2}\rangle\otimes[(|3_{1}0_{2}\rangle-|0_{1}3_{2}\rangle)
\cr&&-\sqrt{3}(|2_{1}1_{2}\rangle-|1_{1}2_{2}\rangle)].\label{e21}
\end{eqnarray}
The small atom-field interaction breaks this four-fold degeneracy,
leading to a unique state with the lowest energy. As shown in Fig.
5(d), for large hopping this eigenstate can be identified as a
polaritonic superfluid state when $h\simeq-\Delta$. The probability
distributions of such two eigenstates with respect to the eigenspace
$\Gamma_k$ and atomic excitation number $N_A$ are plotted in Fig. 6.
In this case all the polariton subspaces are occupied and atoms are
partially excited, indicating that the superfluid state has
polaritonic characteristics.

\begin{figure}
\center
  \includegraphics[width=1\columnwidth]{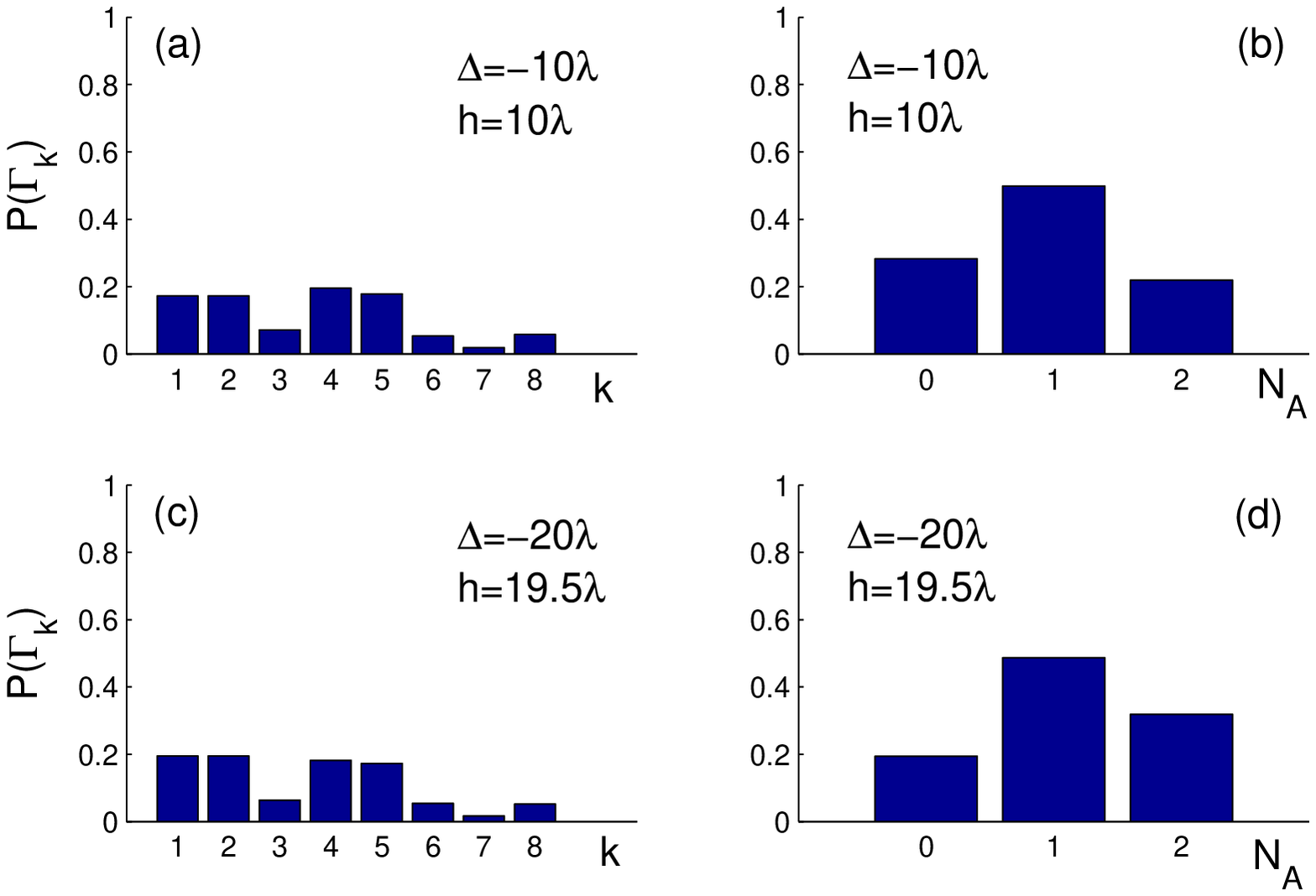} \caption{(Color
  online) [(a)-(c)] Probability distribution $P(\Gamma_{k})$ [$k=1,2,...,8,$
  corresponding to Eqs. (8)-(15)]
  of the lowest energy state with respect to the eigenspace of the
  Hamiltonian without hopping. [(b)-(d)] Probability distribution of the
  total atomic excitation number $N_A$.
   }\label{Fig.6.}
\end{figure}

\section{Total excitation number $N$}

In this section, we examine the dependence of the variances $\Delta
N_1/N$, $\Delta N_{1A}$, and $(\Delta N_1/N)\Delta N_{1A}$ on the
total excitation number $N$ in the lowest energy state with fixed
hopping strength. Here, the relative total excitation number
variance $\Delta N_1/N$ is used to eliminate the effect of simply
expanding the total excitation number and characterize the
polaritonic superfluid.(see Sec. IV.B. for discussion).

\subsection{Small hopping}

\begin{figure}
\center
  \includegraphics[width=0.8\columnwidth]{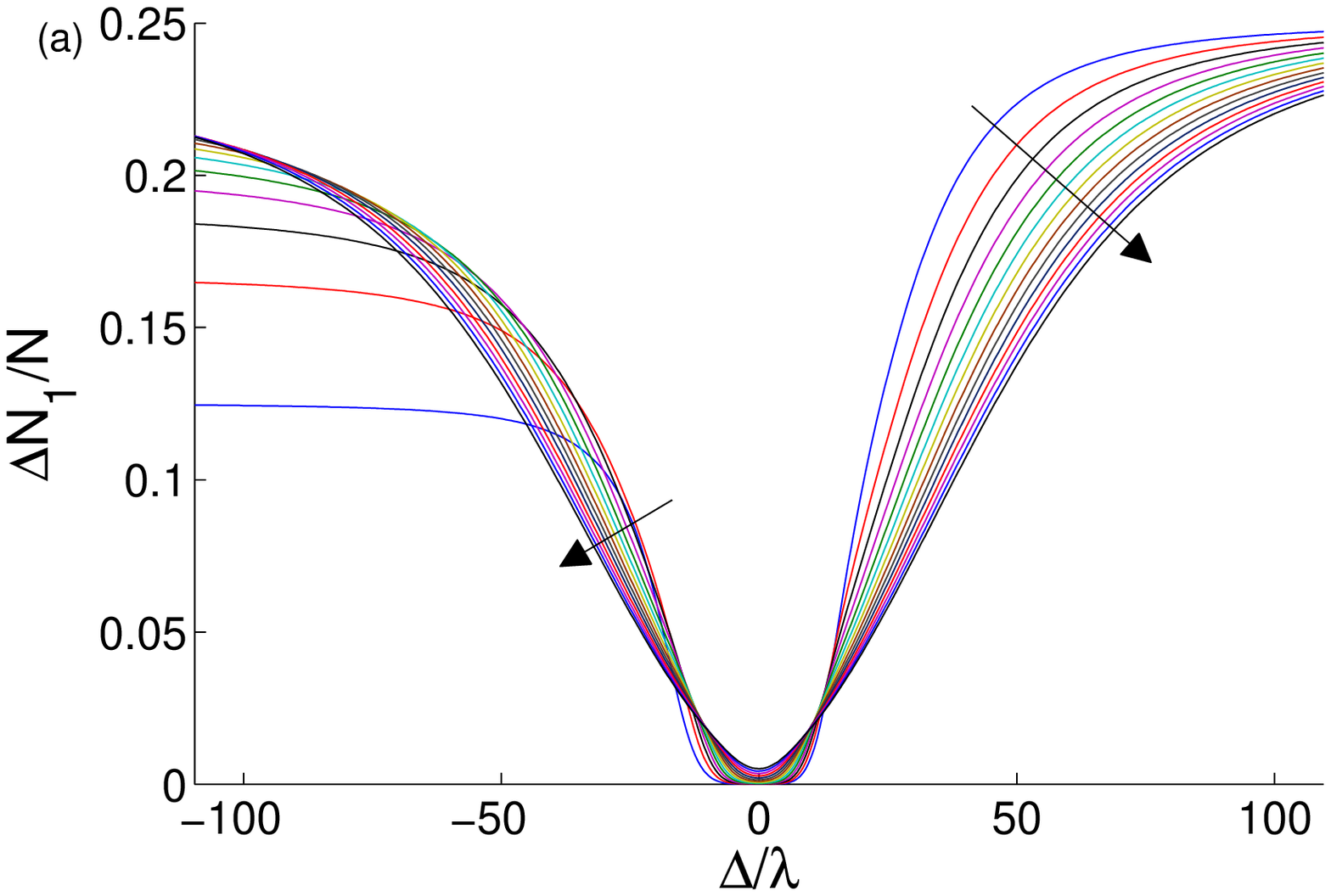}
  \includegraphics[width=0.8\columnwidth]{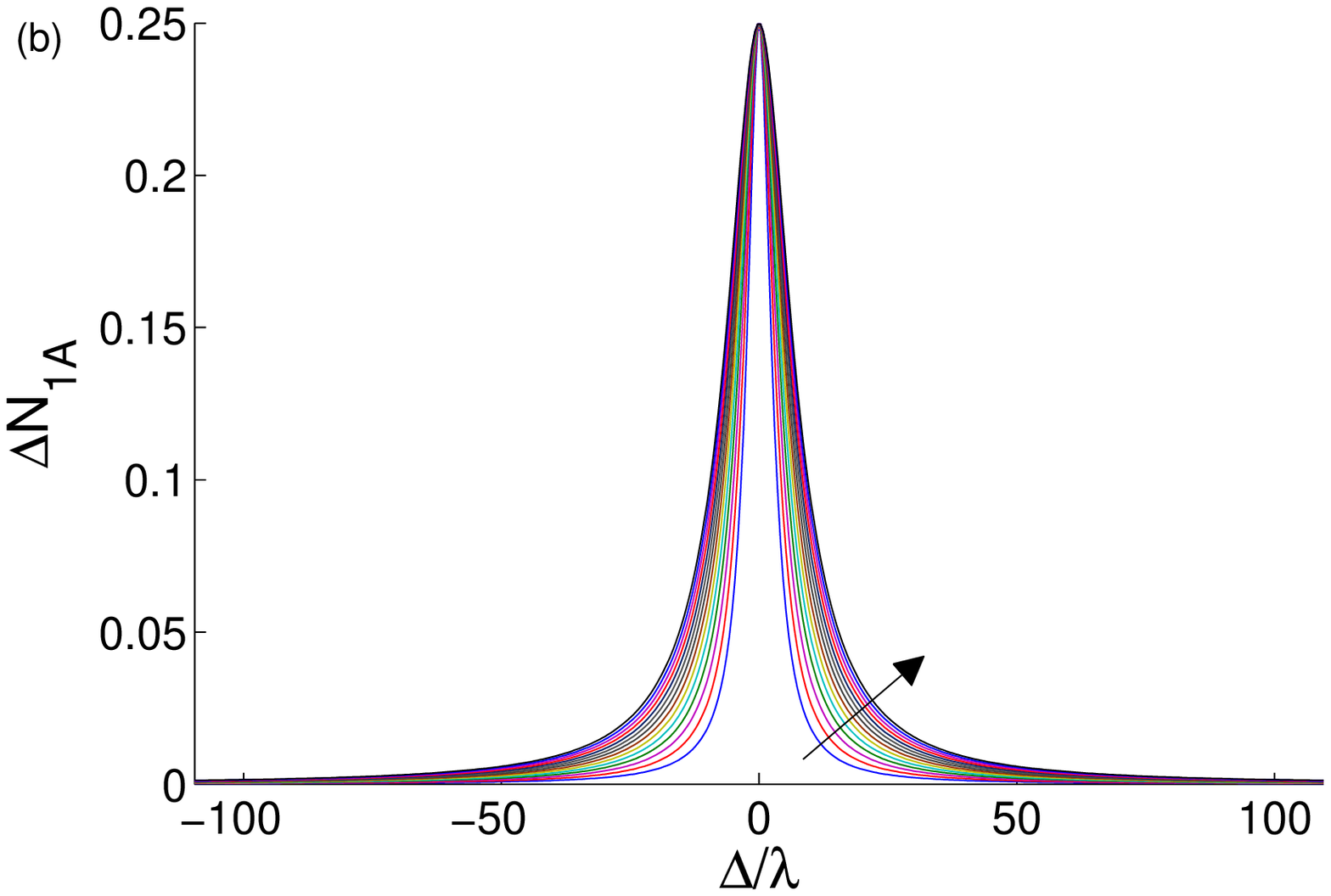}
  \includegraphics[width=0.8\columnwidth]{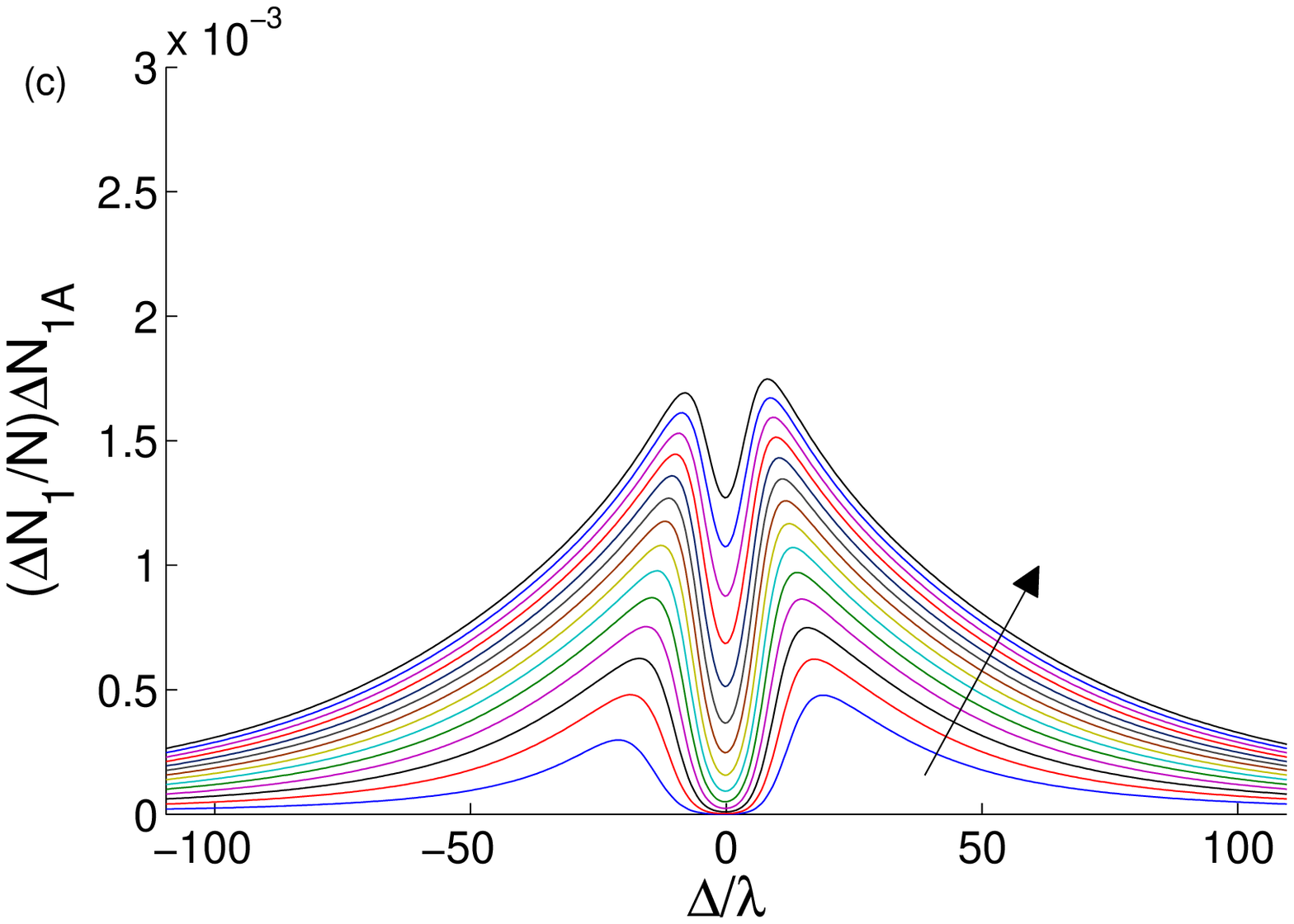} \caption{(Color
  online) Excitation number variance as a function of the
  detuning $\Delta$ in the lowest energy state of the $N$-excitation ($N=4,6,8,...,30$)
  coupled atom-cavity system when $h=10^{-4}\lambda$, where different excitation
  number variances are:
  (a) $\Delta N_{1}/N$; (b) $\Delta N_{1A}$; (c) $(\Delta N_{1}/N)\Delta
  N_{1A}$. The direction of the black arrows represents the
  increasing trend for $N$ in these subfigures.
  }\label{Fig.7.}
\end{figure}

In Fig. 7, we plot different excitation number variances as
functions of the detuning in the lowest energy state of the
$N$-excitation coupled atom-cavity system under the small-hopping
situation, where $N=4,6,8,...,30$.

Figure 7(a) shows the effect of increasing $N$ on the relative total
excitation number variance $\Delta N_1/N$. As before, $\Delta N_1/N
> 0$ indicates a delocalized or superfluidlike state, while $\Delta
N_1/N \sim 0$ indicates a localized, insulatorlike state. For large
positive detuning, $\Delta N_1/N > 0$ corresponds to the photonic
superfluid state, while for large negative detuning, $\Delta N_1/N >
0$ corresponds to the coexisting state with characteristics of both
photonic superfluid and atomic insulator. In the limit of very large
$|\Delta|$, as $N$ increases, the relative excitation number
variance decreases, and the transition from the insulator to
superfluid becomes slower. Figure 7(a) also shows that the region
over which $\Delta N_1/N \sim 0$, indicating an insulator state,
narrows as $N$ increases. The photon hopping strength is set to be
$h=10^{-4}\lambda$ similar to that of Sec. III.A. For arbitrary even
$N>2$ the gap between the lowest two states for $h=0$ becomes
$(2\sqrt{N}-\sqrt{N-1}-\sqrt{N+1})\lambda$, which goes to zero as
$N\rightarrow\infty$. Therefore as $N$ increases the photon
blockade, which leads to the polaritonic insulator state, weakens
and hence can be easily overcome by the photon hopping. When
$N\geq20$, the system can not stay in the insulator state for
$h=10^{-4}\lambda$.

Figure 7(b) shows the atomic excitation number variance $\Delta
N_{1A}$ as a function of $\Delta/\lambda$ for different $N$. As $N$
increases, the maximum value of $\Delta N_{1A}$ remains unchanged,
but the region with nonzero $\Delta N_{1A}$ is broadened. The
maximum variance of the atomic excitation number depends on the
number of atoms in the system, which is independent of the
excitation number $N$. However, the on-resonance Rabi frequency
scales as $\sqrt{N} \lambda$. Therefore the range of detunings over
which the atom-field interaction is large enough to produce
polaritonic behavior increases with $N$.

The combined effect of these features is shown in Fig. 7(c), where
the product of the two variances as a function of $\Delta/\lambda$
is plotted. Here a nonzero value corresponds directly to a
polaritonic superfluid state. It is clear that the polaritonic
superfluid region extends over a wider range of $\Delta$ values as
$N$ increases. The dip at $\Delta = 0$ that indicates the transition
to a polaritonic insulator state fails to go to $0$ when $N\geq20$,
because, as discussed above, the value of $h$ used in this plot is large
 enough to overcome the energy gap between the lowest two levels produced
 by the atom-cavity interaction when $N\geq20$.

\begin{figure}
\center
  \includegraphics[width=0.8\columnwidth]{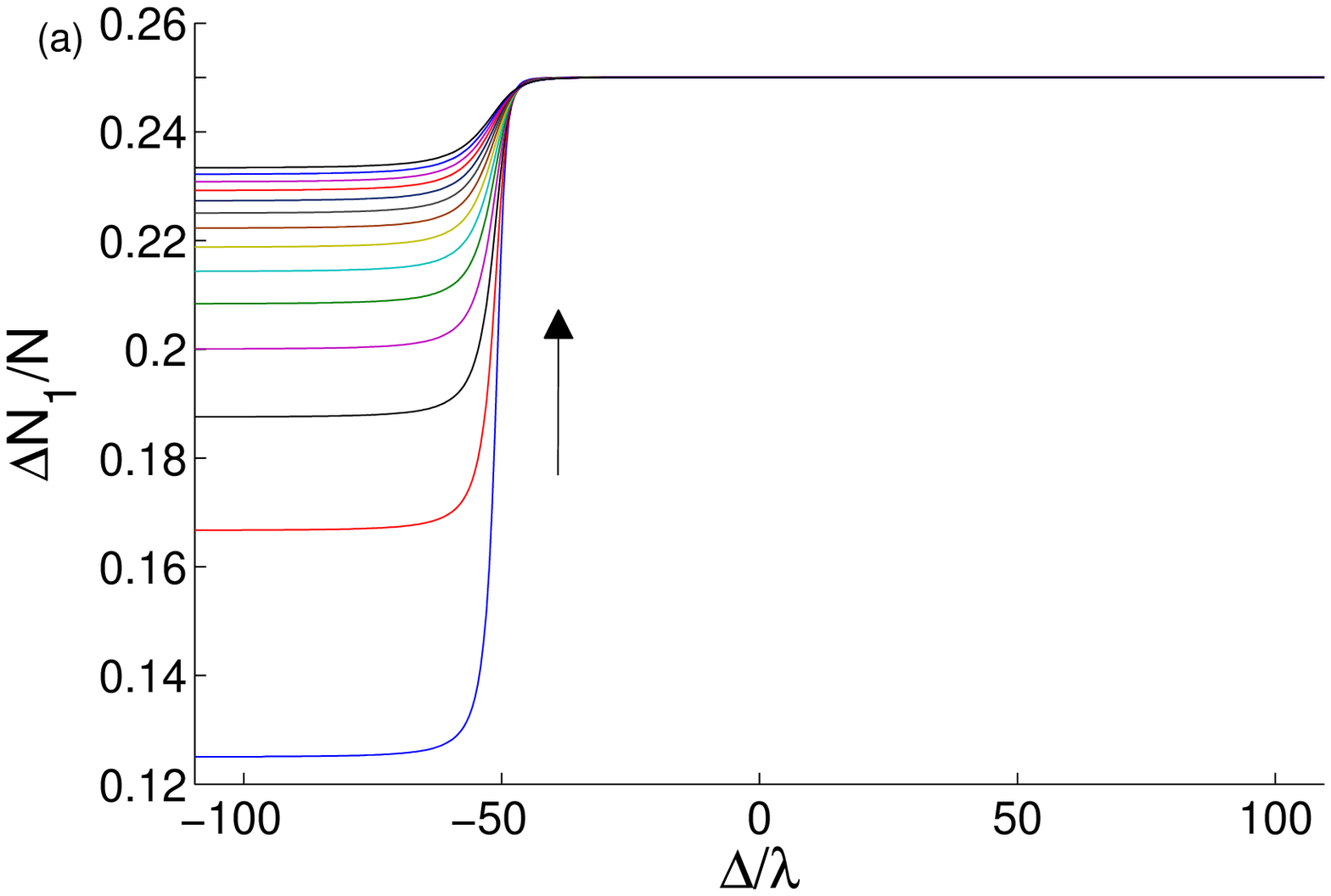}
  \includegraphics[width=0.8\columnwidth]{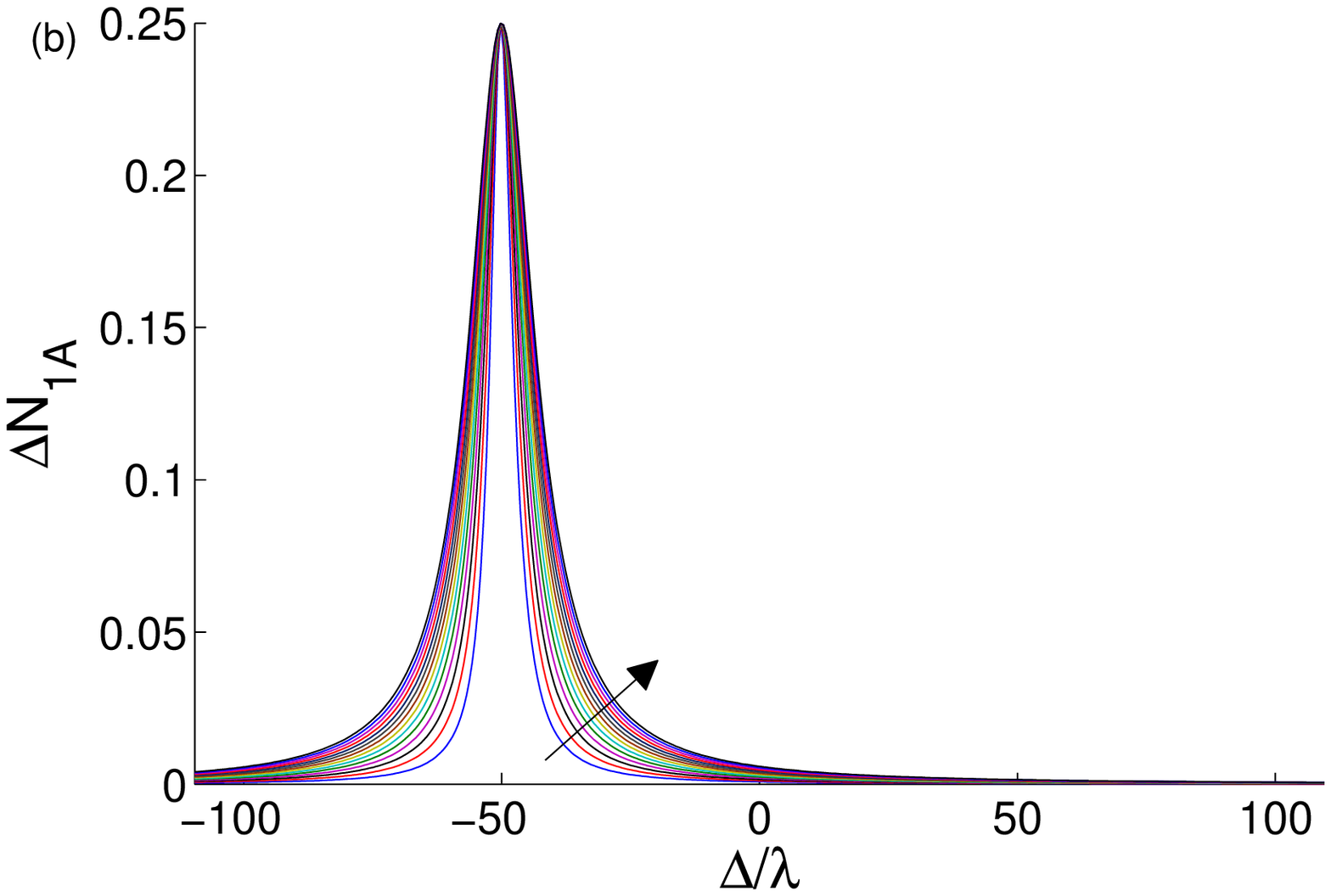}
  \includegraphics[width=0.8\columnwidth]{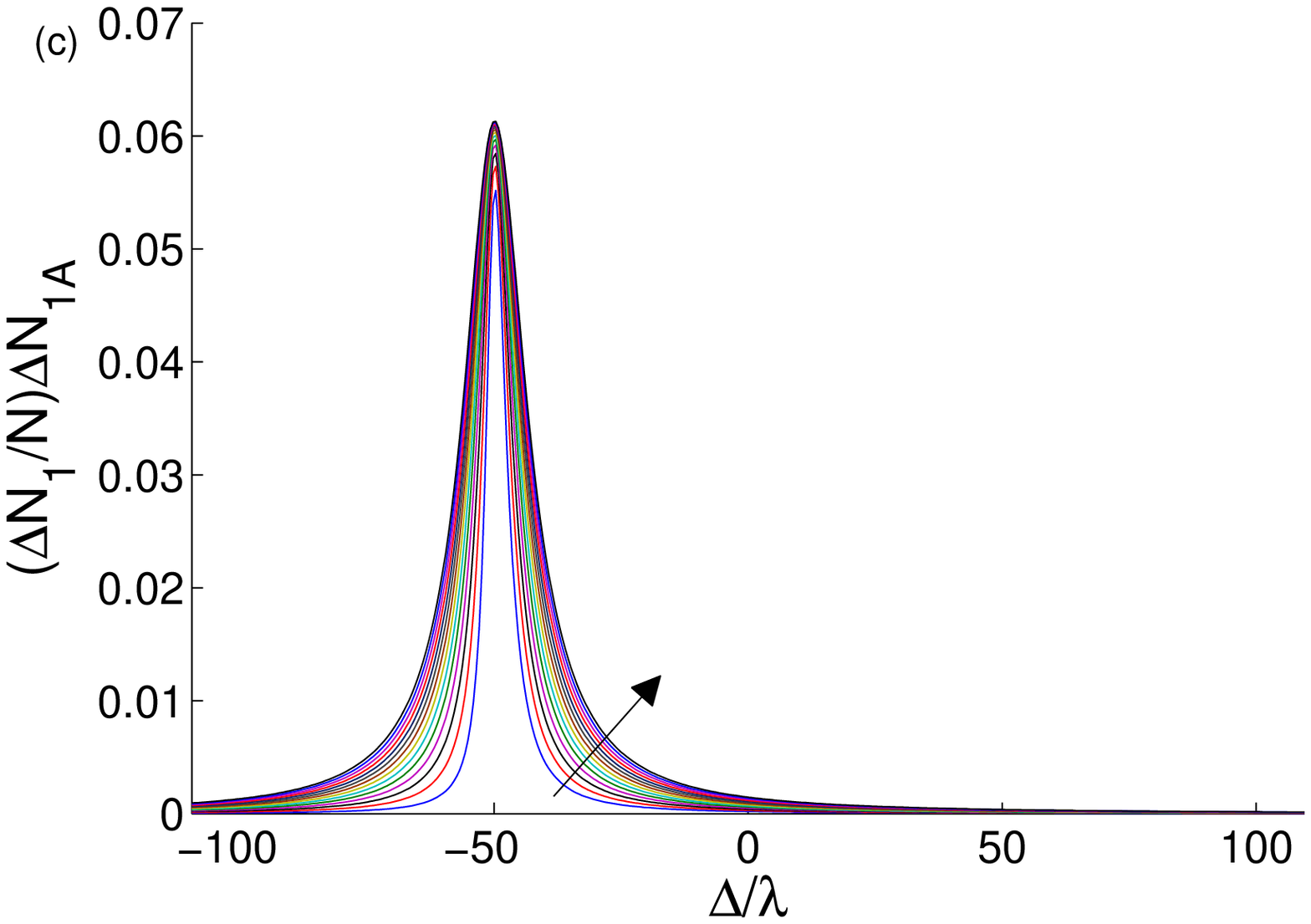} \caption{(Color
  online) Excitation number variance as a function of the
  detuning $\Delta$ in the lowest energy state of the $N$-excitation ($N=4,6,8,...,30$)
  coupled atom-cavity system when $h=50\lambda$, where different excitation number variances are:
  (a) $\Delta N_{1}/N$; (b) $\Delta N_{1A}$; (c) $(\Delta N_{1}/N)\Delta
  N_{1A}$. The direction of the black arrows represents the
  increasing trend for $N$ in these subfigures.
   }\label{Fig.8.}
\end{figure}

\subsection{Small atom-field interaction}

In Fig. 8, we plot various excitation number variances as functions
of the detuning for the lowest energy state of the $N$-excitation
coupled atom-cavity system, where $N=4,6,8,...,30$, in the small
atom-field interaction regime.

Figure 8(a) shows that, in the region $\Delta\leq-h$, the relative
excitation number variance $\Delta N_{1}/N$ becomes larger as $N$
increases, indicating that the superfluid is enhanced. This is due
to the fact that when $\Delta$ passes the critical point
$\Delta_{c}=-h$, the values of $\Delta N_1/N$ all converge to the
maximum $1/4$. Fig. 8(b) shows the atomic excitation number variance
$\Delta N_{1A}$ as a function of $\Delta$ for different $N$ when
$h=50\lambda$. The result is similar to that of Fig. 7(b), except
that the nonzero region indicating that polaritonic states is now
centered around $\Delta=-h$. The product of the two variances as a
function of $\Delta/\lambda$, plotted in Fig. 8(c), demonstrates the
existence of a distinct polaritonic superfluid state in the vicinity
of $\Delta = -h$. As $N$ increases, the width of the polaritonic
superfluid region also increases. The maximum value of the variance
product also increases with $N$. These features suggest that, in the
large hopping limit, the polaritonic superfluid state may be easier
to observe as the total excitation number increases.

\begin{figure}
\center
  \includegraphics[width=0.8\columnwidth]{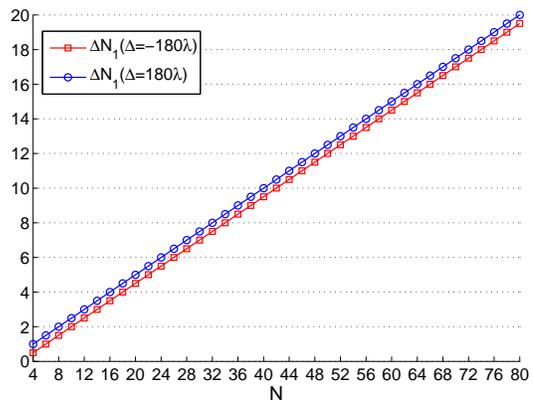}\caption{(Color
  online) Excitation number variance $\Delta N_1$ as a function of $N$ in the lowest
  energy state for the large-detuning limits, where $h=25\lambda$.
   }\label{Fig.9.}
\end{figure}

In Fig. 9, we plot the variance of the total excitation number in
the first site $\Delta N_1$ as a function of the total excitation
number for $\Delta\rightarrow-\infty$ and $\Delta\rightarrow\infty$.
The results show that $\Delta N_1$ has a linear dependence on $N$ in
both limits. We explain this result as follows. In the limit of
large detuning, the two field modes effectively decouple from the
atoms and can be approximately described by the interaction
Hamiltonian:
\begin{eqnarray}
H_{ph}&=&w_c(a_{1}^{\dagger}a_{1}+a_{2}^{\dagger}a_{2})
+h(a_{1}^{\dagger}a_{2}+a_{2}^{\dagger}a_{1}).\label{e22}
\end{eqnarray}
$H_{ph}$ is easily diagonalized by defining the delocalized mode
operators:
\begin{eqnarray}
b_{\pm}&=&\frac{1}{\sqrt{2}}(a_{1}\pm a_{2}),\label{e23}
\end{eqnarray}
where $[b_{+},b_{-}]=0$. In terms of these new operators, $H_{ph}$
becomes:
\begin{eqnarray}
H_{ph}&=&(w_c+h)b_{+}^{\dagger}b_{+}+(w_c-h)b_{-}^{\dagger}b_{-}.\label{e24}
\end{eqnarray}
Assuming $h>0$, the lowest energy state in its $N$-excitation
subspace is given by:
\begin{eqnarray}
|\psi_{g,N}^{ph}\rangle&=&\frac{1}{\sqrt{N!}}(b_{-}^{\dagger})^{N}|0_{1}\rangle|0_{2}\rangle.\label{e25}
\end{eqnarray}
Thus, we straightforwardly calculate $\Delta N_1$ in the
large-detuning limits:
\begin{eqnarray}
\Delta N_1&\simeq&\frac{(N-2)}{4}, (\Delta\rightarrow-\infty)\label{e26}\\
\cr \Delta N_1&\simeq&\frac{N}{4},
(\Delta\rightarrow\infty)\label{e27}
\end{eqnarray}
which analytically demonstrates the linear relationship between
$\Delta N_1$ and $N$ in the large-detuning limit.

For large positive $\Delta$, the atoms are both in the ground states
and so the lowest energy state of the system is given by
$|\psi_{g,N}^{ph}\rangle$: all $N$ excitations go into the
delocalized mode $b_{-}$. On the other hand, for large negative
$\Delta$ the atoms are both excited, leaving only $N-2$ excitations
in the field modes. In other words, in this regime the lowest energy
state is the combination of the atomic insulator state and the
photonic superfluid state. The effect of the atomic insulator state
on the total excitation variance is weakened as $N$ increases. This
behavior is evident in Fig. 7(a) and Fig. 8(a). The state
$|\psi_{g,N}^{ph}\rangle$ has a larger number variance for large $N$
but is not in a sense any more delocalized, since the delocalized
mode defined by $b_{-}$ is maximally delocalized over the two sites.
Larger $N$ simply means there are more photons in the delocalized
mode $b_{-}$, which explains the behavior shown in Fig. 8(a), where
the curves for different $N$ all converge to the same constant.

\section{Conclusion}

In summary, we have investigated the quantum phase transition
behavior of polaritonic excitations in a multi-excitation coupled
atom-cavity system. By examining our system with various parameters,
we have identified different phases in the lowest energy state. The
case of total excitation number $N=4$ has been treated both
analytically and numerically, and the results are then generalized
to the case with higher excitation numbers. In the small
photon-hopping and the small atom-field interaction cases, we have
identified an interesting coexisting phase involving characteristics
of both photonic superfluid and atomic insulator. We find that the
region where the system exhibits the polaritonic characteristic
becomes broader as the total excitation number increases. Finally,
we demonstrate that $\Delta N_1$ has a linear dependence on $N$ in
the large-detuning limit. The results we have presented are not
limited to the cavity QED system, and are general and applicable to
all the systems, such as ion traps and circuit QED systems.

\section{Acknowledgement}

We would like to thank Dr. Kenji Toyoda for useful discussions. EKI
acknowledges funding from the Leverhulme Trust. This work is
supported by the Major State Basic Research Development Program of
China under Grant No. 2012CB921601, the National Natural Science
Foundation of China under Grant No. 11374054, No. 11305037, No.
11347114, and No. 11247283, the Natural Science Foundation of Fujian
Province under Grant No. 2013J01012, and  the funds from Fuzhou
University under Grant No. 022513, Grant No. 022408, and Grant No.
600891.

\end{document}